\documentclass[prd,amsmath,twocolumn,10pt,superscriptaddress,floatfix,nofootinbib]{revtex4-2}
\usepackage{epsfig,amssymb,amsfonts,amsmath,mathtools,bm,color,xcolor,graphicx,braket,adjustbox,esint,upgreek}
\usepackage{multirow}
\usepackage[utf8]{inputenc}
\usepackage{setspace}
\usepackage{dcolumn,kantlipsum}
\allowdisplaybreaks

\usepackage[
colorlinks=true,
linkcolor=blue,
breaklinks=true,
urlcolor=blue,
citecolor=blue]{hyperref}

\usepackage{orcidlink}
\usepackage{mathrsfs} 

\usepackage{bm,bbm}
\usepackage{slashed}

\newcommand{\be}{\begin{equation}}
\newcommand{\ee}{\end{equation}}
\newcommand{\bea}{\begin{eqnarray}}
\newcommand{\eea}{\end{eqnarray}}
\newcommand{\beas}{\begin{eqnarray*}}
\newcommand{\eeas}{\end{eqnarray*}}

\renewcommand{\arraystretch}{1.2}
\def\vec#1{\boldsymbol{#1}}

\renewcommand{\L}{\mathscr{L}}
\newcommand{\M}{\mathscr{M}}
\newcommand{\MeV}{\,\text{MeV}}
\newcommand{\GeV}{\,\text{GeV}}
\renewcommand{\vec}[1]{\mathbf{#1}}
\newcommand{\diff }{{\text{d}}}

\newcommand{\zc}{Z_c(3900)}

\newcommand{\ustb}{\affiliation{School of Mathematics and Physics, University
of Science and Technology Beijing, Beijing 100083, China}}

\newcommand{\uestc}{\affiliation{School of Physics, University of Electronic Science and Technology of China, Chengdu 611731, China}}

\newcommand{\itp}{\affiliation{CAS Key Laboratory of Theoretical Physics, Institute of Theoretical Physics,\\ Chinese Academy of Sciences, Beijing 100190, China}}

\newcommand{\ucas}{\affiliation{School of Physical Sciences, University of Chinese Academy of Sciences, Beijing 100049, China}}

\newcommand{\peng}{\affiliation{Peng Huanwu Collaborative Center for Research and Education, Beihang University, Beijing 100191, China}}

\graphicspath{{FiguresV2/}}

\begin{document}

\title{Precise determination of the pole position of the exotic $Z_c(3900)$}

\author{Yun-Hua Chen\orcidlink{0000-0001-8366-2170}}\email{Corresponding author, yhchen@ustb.edu.cn}
\ustb

\author{Meng-Lin Du\orcidlink{0000-0002-7504-3107}}\email{Corresponding author, du.ml@uestc.edu.cn}
\uestc

\author{Feng-Kun Guo\orcidlink{0000-0002-2919-2064}}\email{Corresponding author, fkguo@itp.ac.cn}
\itp \ucas \peng

% \begin{onecolumngrid}

\begin{abstract}

Abstract:  We perform a unified description of the experimental data of
the $\pi^+\pi^-$ and
$J/\psi\pi^\pm$ invariant mass spectra for $e^+e^-
\rightarrow J/\psi \pi^+\pi^-$ and the $D^0 D^{\ast-}$ mass spectrum for $e^+e^-
\rightarrow D^0 D^{\ast-} \pi^+$ at $e^+e^-$ center-of-mass energies 4.23 and 4.26~GeV.
The analysis takes into account open-charm meson loops that contain triangle singularities, the $J/\psi\pi$-$D\bar D^*$ coupled-channel interaction respecting unitarity,
and the strong $\pi\pi$-$K\bar K$ final state interaction using dispersion relations.
The analysis leads to a precise determination of the $Z_c(3900)$ pole with the pole mass and width $(3880.7 \pm 1.7_\text{stat}\pm 22.4_\text{syst})$ MeV and $(35.9 \pm 1.4_\text{stat}\pm 15.3_\text{syst})$ MeV, respectively, and hints at that the $D\bar D^*$ molecular and non-molecular components are of similar importance in the $Z_c(3900)$ formation.

\medskip



\end{abstract}
% \end{onecolumngrid}

\maketitle
\medskip
\medskip
\medskip

\textbf{1. Introduction}\\

The nature of the charged charmoniumlike $Z_c(3900)^\pm$ structure is still in heated debate although a decade has passed since its discovery
by the BESIII and Belle Collaborations in 2013 in the process $e^+e^-\to  J/\psi\pi^+\pi^-$~\cite{Ablikim:2013mio,Liu:2013dau}. It was subsequently confirmed by an analysis of the CLEO-c data~\cite{Xiao:2013iha} and from the semi-inclusive $b$-hadron decays at the D0 experiment~\cite{D0:2018wyb}.
It couples much stronger to $D\bar D^*$ than to $J/\psi\pi$ given the ratio of partial widths extracted by BESIII: $\Gamma((D\bar D^*)^\pm)/\Gamma(J/\psi\pi^\pm)=6.2\pm1.1\pm2.7$~\cite{BESIII:2013qmu}.
Being explicitly charged in the charmonium mass regime, the $Z_c(3900)^\pm$ necessarily bears at least four valence quarks if it is a genuine hadronic resonance, an $S$-matrix pole independent of the reaction kinematics. It was proposed to be a $D\bar D^*$ hadronic molecule~\cite{Wang:2013cya,Guo:2013sya, Wilbring:2013cha, He:2013nwa,Dong:2013iqa,Zhang:2013aoa, Aceti:2014uea, Albaladejo:2015lob,Albaladejo:2016jsg, Gong:2016hlt,Gong:2016jzb,He:2017lhy,Ortega:2018cnm, Du:2020vwb,Meng:2020ihj,Wang:2020dgr} or a compact tetraquark state~\cite{Braaten:2013boa,Dias:2013xfa,Maiani:2014aja,Qiao:2013raa,Deng:2014gqa,Agaev:2017tzv}.
Meanwhile, the proximity of its mass, $(3887.1\pm2.6)$~MeV, the value quoted in~\cite{ParticleDataGroup:2022pth} obtained by averaging those from experimental analyses of both charged and neutral $Z_c(3900)$~\cite{Liu:2013dau,Xiao:2013iha,BESIII:2013qmu,BESIII:2015pqw,BESIII:2015ntl,BESIII:2017bua,D0:2019zpb,BESIII:2020oph}, to
the $ D\bar{D}^\ast$ threshold also brings up interpretations of possible kinematical effects, such as threshold cusp due to coupled channels without a pole~\cite{Chen:2013wca,Swanson:2014tra,HALQCD:2016ofq} or that triangle singularity (TS) may play an important role~\cite{Wang:2013cya,Szczepaniak:2015eza,Albaladejo:2015lob,Pilloni:2016obd,vonDetten:2023uja}; see~\cite{Chen:2016qju,Hosaka:2016pey,Lebed:2016hpi,Esposito:2016noz,Guo:2017jvc,Ali:2017jda,Olsen:2017bmm,Karliner:2017qhf,Yuan:2018inv,Liu:2019zoy,Brambilla:2019esw,Guo:2019twa,JPAC:2021rxu,Meng:2022ozq} for recent reviews.

The averaged width in~\cite{ParticleDataGroup:2022pth} is $(28.4\pm2.6)$~MeV.
Most experimental analyses use the Breit-Wigner parameterization~\cite{Ablikim:2013mio,Liu:2013dau,Xiao:2013iha,BESIII:2013qmu,BESIII:2015pqw,BESIII:2015ntl,D0:2018wyb,D0:2019zpb,BESIII:2020oph}. The only exception is~\cite{BESIII:2017bua} with the largest number of events ($\sim6\times 10^3$). It employs the Flatt\'e parameterization and concludes with large systematic uncertainties, coming mainly from the parameterization of the light scalar mesons, for the mass and width: $(3881.2\pm4.2_\text{stat.}\pm52.7_\text{syst.})$~MeV and $(51.8\pm4.6_\text{stat.}\pm36.0_\text{syst.})$~MeV, respectively.

It has been argued~\cite{Guo:2014iya,Dong:2020hxe} that for a prominent near-threshold enhancement, in particular in the elastic channel (that is the channel with the relevant threshold), unitarity requires the amplitude possess a pole not far from the threshold if there are no complications due to other kinds of singularities such as the TS. Namely, a $T$ matrix fulfilling unitarity needs to be considered when describing near-threshold structures; the case of $\zc$ was analyzed in~\cite{Guo:2014iya,Pilloni:2016obd,He:2017lhy,Baru:2021ddn,Du:2022jjv,Yan:2023bwt}.
It was first pointed out in~\cite{Wang:2013cya} that although the singularity due to triangle diagrams with the intermediate $D_1\bar D D^*$ states (here and below $D_1$ and $D_{s1}$ mean $D_1(2420)$ and $D_{s1}(2536)$, respectively) can leave a significant impact on the $Z_c(3900)$ structure, a pole corresponding to the $\zc$ is still needed to get a decent description of the BESIII $J/\psi\pi\pi$ data~\cite{Ablikim:2013mio}.
A similar conclusion was reached in~\cite{Albaladejo:2015lob} by simultaneously analyzing the $D\bar D^*$~\cite{BESIII:2015pqw} and $J/\psi\pi$~\cite{Ablikim:2013mio} invariant mass distributions by considering both the triangle diagrams and $J/\psi\pi$--$D\bar D^*$ coupled-channel $T$ matrix respecting unitarity.
Depending on the parameterizations of the $D\bar D^*$ interaction, the data are consistent with the $\zc$ being either a virtual state below the $D\bar D^*$ threshold or an above-threshold resonance.
The analysis was extended in~\cite{Du:2022jjv} to include the BESIII data of both $\zc$~\cite{BESIII:2015pqw,BESIII:2017bua} and $Z_{cs}(3985)$~\cite{BESIII:2020qkh} with flavor SU(3) constraints, and it was found that the resonance scenario for the $\zc$ is preferred though the virtual state one is not excluded (for similar analyses focusing on the $Z_{cs}(3985)$, see~\cite{Yang:2020nrt,Baru:2021ddn}).
In~\cite{Albaladejo:2015lob,Du:2022jjv}, however, the $\pi\pi$ invariant mass spectrum was not considered, and thus neither was the crossed-channel effects from the $\pi\pi$ final state interaction (FSI) included.
The JPAC Collaboration analyzed the BESIII data of both the charged~\cite{Ablikim:2013mio,BESIII:2015pqw} and neutral~\cite{BESIII:2015cld,BESIII:2015ntl} channels~\cite{Pilloni:2016obd}. The $\pi\pi$ spectrum was fitted through two ``effective'' scalar resonances: $f_0(500)$ and $f_0(980)$ with resonance parameters determined from the fit. Open charm triangles were also considered.
It was found that the data could be described with similar qualities in several scenarios, either with a near-threshold $\zc$ pole or without; for the latter, the nontrivial structure is generated by open-charm TSs.
One difference between~\cite{Albaladejo:2015lob} and~\cite{Pilloni:2016obd} is that the $D_1D^*\pi$ coupling is purely $D$-wave in the former and purely $S$-wave in the latter.
Both $S$- and $D$-wave $D_1D^*\pi$ couplings are considered in~\cite{Du:2022jjv} following~\cite{Guo:2020oqk}.
The first fit to the $e^+e^-\to J/\psi\pi^+\pi^-$ data with the $\pi\pi$-$K\bar K$ coupled channel FSI considered using dispersion relations was performed in~\cite{Danilkin:2020kce}.
However, the open-charm channels, as well as the open-charm triangle diagrams, were not considered; instead, the $\zc$ propagator was built in explicitly (see also~\cite{vonDetten:2023uja}).

In this article, we give the first unified description of the experimental data of
the $\pi^+\pi^-$ and
$J/\psi\pi^\pm$ invariant mass distributions for $e^+e^-
\rightarrow J/\psi \pi^+\pi^-$~\cite{BESIII:2017bua} and the $D^0 D^{\ast-}$ mass spectrum for $e^+e^-
\rightarrow D^0 D^{\ast-} \pi^+$~\cite{BESIII:2015pqw} at the $e^+e^-$ center-of-mass (c.m.) energies $E=4.23$~GeV and $4.26$~GeV measured by BESIII with dispersive amplitudes by including both open-charm triangles and $\pi\pi$-$K\bar K$ FSI.
The analysis leads to a precise determination of the $Z_c(3900)^\pm$ pole.\\

\textbf{2. Framework} \\

\begin{figure}[tb]
\centering
\includegraphics[width=\linewidth]{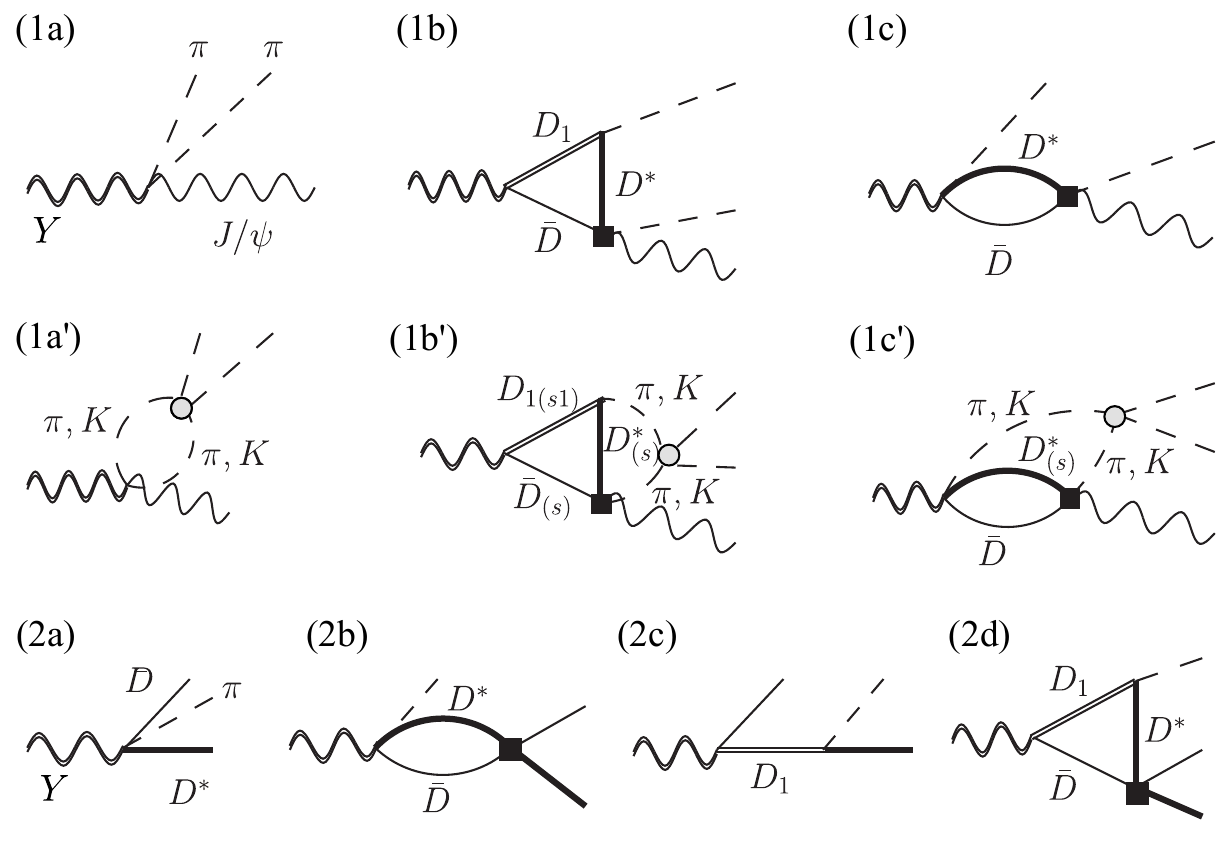}
  \caption{Feynman diagrams for $Y \to J/\psi \pi \pi $ and $Y \to D^{\ast}\bar{D} \pi $. For $Y \to J/\psi \pi \pi$ process: (1a) and ($\text{1a}^\prime$) denote  the contributions of the chiral contact $Y\psi\Phi\Phi$ terms; (1b) and ($\text{1b}^\prime$) correspond to the triangle diagrams; (1c) and ($\text{1c}^\prime$) denote the bubble diagrams.
  For $Y \to D^{\ast}\bar{D} \pi$ process, (2a), (2b), (2c), and (2d) denote the contribution of the chiral contact $Y D^\ast\bar{D}\phi$ term, the bubble diagram, the $D_1$-exchange term, and the triangle diagram, respectively.
  The double-wavy line, wavy line, and dashed lines denote the $Y$, $J/\psi$, and $\pi(K)$ mesons, respectively. The solid, thick solid, and double-solid lines denote the $\bar D_{(s)}$, $D_{(s)}^*$, $D_{1(s1)}$ mesons, respectively.
  The gray blob denotes the $\pi\pi$-$K\bar{K}$ FSI and the filled square denotes the $J/\psi\pi(J/\psi \bar{K})$-$\bar{D}D^*(\bar{D}D^*_s)$ FSI.
  The charge conjugated charm loops are not shown but are included in the calculation.
    }
  \label{fig.FeynmanDiagram}
\end{figure}

We denote the source of $J/\psi\pi\pi$ and $D\bar D^*\pi$ as $Y$, which couples to the virtual photon from the $e^+e^-$ annihilation.
We consider the diagrams shown in Fig.~\ref{fig.FeynmanDiagram}. The triangle diagrams $Y \rightarrow \bar{D}D_1/\bar{D}_s D_{s1}\rightarrow \bar{D}D^\ast\pi (\bar{D}D_s^\ast K) / \bar{D}_s D^\ast K $~\cite{Cleven:2013mka,Wang:2013hga,Albaladejo:2015lob,Guo:2020oqk,Du:2022jjv} contain TS contributions, and the $\bar{D}D^\ast (\bar{D}D_s^\ast )$ loops followed by the $J/\psi\pi(J/\psi \bar{K})$-$\bar{D}D^*(\bar{D}D^*_s)$ coupled-channel rescattering allows for the existence of a $Z_c (Z_{cs})$ pole.
The $S$-wave $\pi\pi$-$K\bar K$ coupled-channel FSI\footnote{We need to consider the $ e^+e^- \to J/\psi K \bar K $  process in the $\pi\pi$-$K\bar{K}$ FSI.} and $D$-wave $\pi\pi$ FSI will be treated model-independently using the dispersion approach based on unitarity and analyticity---the modified Omn\`es representation of the Khuri-Treiman type~\cite{Anisovich:1996tx}.

The double differential decay width for the process $Y(p_a) \rightarrow J/\psi(p_b)\pi(p_c)\pi(p_d)$ can be written as
\begin{align}
\frac{\diff^2 \Gamma}{\diff s \diff t}    = \frac{1}{32(2\pi)^3 M_Y^3}\frac{1}{3}\sum_{\lambda, \lambda^\prime} \left|\mathscr{M}^{\lambda,\lambda^\prime}(s,t,u)\right|^2\,,
\label{eq.DifDecayWidthJpsipipi}
\end{align}
where $M_Y=E$ is the $e^+e^-$ c.m. energy, and the Mandelstam variables are
$s=(p_c+p_d)^2$, $t=(p_a-p_c)^2$, and $u=(p_a-p_d)^2$. $\mathscr{M}^{\lambda,\lambda^\prime}(s,t,u)\equiv \mathscr{M}_{\mu\nu}(s,t,u)\epsilon_{\lambda}^\mu \epsilon_{\lambda^\prime}^{\ast\nu}$ is helicity amplitude, and $\lambda$ and $\lambda^{\prime}$ denote the helicity of $Y$ and $J/\psi$, respectively. To account for the strong $\pi\pi$ FSI using dispersion theory, we perform the $s$-channel partial-wave decomposition of the amplitude
\begin{align}\label{eq.PWHelicityAmp}
\mathscr{M}^{\lambda,\lambda^\prime}(s,t,u)  = \sum_{l=0}^{\infty}  h_\pi^{\lambda, \lambda^\prime,l}(s) d_{\lambda-\lambda^\prime,0}^l(\theta)\,,
\end{align}
where $l$ represents the relative orbital angular momentum of the pions, $\theta$ denotes the angle between $\pi^+$ and $Y$ in the $\pi\pi$ c.m. frame, and $d_{\lambda-\lambda^\prime,0}^l(\theta)$ is the Wigner-$d$ function. Parity and $C$-parity conservations require $l$ to be even.
The partial-wave amplitude including the $\pi\pi$ FSI reads
\begin{align}
h_\pi^{\lambda, \lambda^\prime,l}(s)  = H_\pi^{\lambda, \lambda^\prime,l}(s)+\hat{H}_\pi^{\lambda, \lambda^\prime,l}(s)\,, \label{eq:h}
\end{align}
where $H_\pi^{\lambda, \lambda^\prime,l}(s)$ is
the right-hand cut (rhc) part and represents the $s$-channel $\pi\pi$ rescattering. The ``hat function'' $\hat{H}_\pi^{\lambda, \lambda^\prime,l}(s)$ encodes the left-hand cut (lhc) contribution of the crossed-channel effects from Fig.~\ref{fig.FeynmanDiagram} (1b) and (1c). Unitarity implies that below the inelastic threshold, the discontinuity of $H(s)$ by crossing the unitarity cut is
\begin{align}\label{eq:disc}
\text{Disc}H_\pi^{\lambda, \lambda^\prime,l}(s) = T^{*}_{\pi,l}(s)\sigma_\pi(s)
h_\pi^{\lambda, \lambda^\prime,l}(s) \,,
\end{align}
where $T_{\pi,l}(s)$ is the corresponding $l$-wave elastic $\pi\pi$ scattering amplitude, and $\sigma_P(s) \equiv
\sqrt{1-4m_P^2/s} $ is the two-body phase space factor.
While we only consider the $S$- and $D$-wave $\pi\pi$ $s$-channel rescattering, we keep the crossed-channel partial-wave expansion to all orders to preserve the singularities in the crossed channels. Namely, in the summation of $l$ in Eq.~\eqref{eq.PWHelicityAmp}, for $H_\pi^{\lambda, \lambda^\prime,l}(s)$ we only consider the $l=0,2$ terms, while for $\hat{H}_\pi^{\lambda, \lambda^\prime,l}(s)$ we take into account all possible $l$ terms~\cite{Danilkin:2020kce,Baru:2020ywb}, i.e., Eq.~\eqref{eq.PWHelicityAmp} is reduced to
\begin{equation}\label{eq.PWHelicityAmp2}
\mathscr{M}^{\lambda,\lambda^\prime}\!(s,t,u)  = \sum_{l=0,2}  H_\pi^{\lambda, \lambda^\prime,l}(s) d_{\lambda-\lambda^\prime,0}^l(\theta) + \hat{\mathscr{M}}(s,\theta)\,.
\end{equation}
This treatment is crucial to make a full Dalitz plot analysis, i.e., to describe the peak structures in both
the $\pi\pi$ and $J/\psi\pi$ invariant spectra.

For the $D$-wave component ($l=2$), the dispersive solution to Eq.~\eqref{eq:disc} is~\cite{Anisovich:1996tx}
\begin{align}\label{OmnesSolution1channel}
H_\pi^{\lambda, \lambda^\prime,2}(s)=&\,\Omega_2^0(s)\bigg\{P_{2,n-1}(s) \\
& +\frac{s^n}{\pi}\int_{4m_\pi^2}^\infty
\frac{\diff x}{x^n}\frac{\hat{H}_\pi^{\lambda, \lambda^\prime,2}(s)\sin\delta_2^0(x)}{|\Omega_2^0(x)|(x-s)}\bigg\} \,, \nonumber
\end{align}
where $P_{2,n-1}(s)$ is a subtraction function, and the single-channel
Omn\`es function reads~\cite{Omnes:1958hv}
\begin{equation}\label{Omnesrepresentation}
\Omega_2^0(s)=\exp
\bigg\{\frac{s}{\pi}\int^\infty_{4m_\pi^2}\frac{\diff x}{x}
\frac{\delta_2^0(x)}{x-s}\bigg\}\,,
\end{equation}
with $\delta_2^0(s)$ the $\pi\pi$ isoscalar $D$-wave phase shift~\cite{Garcia-Martin:2011iqs}.

Since the $\pi\pi$ invariant mass reaches above the $K\bar{K}$ threshold for $Y\to J/\psi\pi^+\pi^-$, for the $S$-wave component one needs to consider the $\pi\pi$-$K\bar K$ coupled-channel FSI. In this way, the amplitude contains both the $f_0(500)$ and $f_0(980)$ contributions.
The discontinuity equation~\eqref{eq:disc} is extended to a two-dimensional matrix form referring to components $\pi\pi$ and $K\bar K$.
Its solution reads
\begin{align}\label{OmnesSolution2channel}
\vec{H}^{\lambda, \lambda^\prime,0}(s)=&\, \Omega(s)\bigg\{\vec{P}_{n-1}(s) \\& +\frac{s^n}{\pi}\int_{4m_\pi^2}^\infty
\frac{\diff
x}{x^n}\frac{\Omega^{-1}(x)T_0^0(x)\Sigma(x)\vec{\hat{H}}^{\lambda, \lambda^\prime,0}(s)}{x-s}\bigg\}
\,, \nonumber
\end{align}
where the two-dimensional vectors
\begin{align}
  \vec{H}^{\lambda, \lambda^\prime,0}(s)&=\big( {H_\pi^{\lambda, \lambda^\prime,0}(s)} ,
  2/\sqrt{3}\,H_K^{\lambda, \lambda^\prime,0}(s)
   \big)^{T}, \notag\\
  \vec{\hat{H}}^{\lambda, \lambda^\prime,0}(s)&=\big( {\hat{H}_\pi^{\lambda, \lambda^\prime,0}(s)} ,
  2/\sqrt{3}\,\hat{H}_K^{\lambda, \lambda^\prime,0}(s)
   \big)^{T}
\end{align}
stand for the rhc and the lhc
parts of the $J/\psi\pi\pi$ and the $J/\psi K\bar{K}$ final states,
respectively, $\Sigma(s)\equiv \text{diag}
\big(\sigma_\pi(s)\theta(s-4m_\pi^2),\sigma_K(s)\theta(s-4m_K^2)\big)$, and $\vec{P}_{n-1}(s)$ is a subtraction function. $\Omega(s)$ is the solution of the homogeneous two-channel $S$-wave discontinuity equation~\cite{Donoghue:1990xh,Moussallam:1999aq,Hoferichter:2012wf,Daub:2015xja}.
The matrix $T_0^0(s)$ reads
\begin{equation}\label{eq.T00}
T_0^0(s)=
 \left( {\begin{array}{*{2}c}
   \frac{\eta_0^0(s)e^{2i\delta_0^0(s)}-1}{2i\sigma_\pi(s)} & |g_0^0(s)|e^{i\psi_0^0(s)}   \\
  |g_0^0(s)|e^{i\psi_0^0(s)} & \frac{\eta_0^0(s)e^{2i\left(\psi_0^0(s)-\delta_0^0(s)\right)}-1}{2i\sigma_K(s)} \\
\end{array}} \right),
\end{equation}
where $\eta_0^0(s)$ is the inelasticity, $\delta_0^0(s)$ is the $\pi\pi$ isoscalar $S$-wave phase shift, $|g_0^0(s)|$ and $\psi_0^0(s)$ are the modulus and phase of the $\pi\pi \to
K\bar{K}$ $S$-wave amplitude, respectively. They are taken from~\cite{Dai:2014lza,Dai:2014zta}.
In this way, the largest systematic uncertainty in the latest BESIII analysis~\cite{BESIII:2017bua}, due to the parametrization of the light scalar mesons, is diminished.

The amplitudes of the charmed-meson loop diagrams $\hat H_{\pi(K)}^{\lambda,\lambda',l}$ involve several parameters: the coupling constant $y$ of the $Y$ to $D_1D$/$D_{s1}D_s$; the coupling constants $h_S$ and $h_D$ of the $S$- and $D$-wave $D_1D^\ast\pi$/$D_{s1}D^\ast K$ vertex, respectively; and the coupling constant $g_{YP}$ of the $Y D^\ast \bar{D}\pi$ vertex. While $h_S$ and $h_D$ have been determined to be $|h_S|=0.57$ and $|h_D|=1.17$ GeV$^{-1}$~\cite{Guo:2020oqk}, $y$ and $g_{YP}$ will be treated as free parameters and determined by fitting the data.

For the contribution from triangle diagrams with the $S$-wave $D_1D^*\pi$ coupling, the subtraction functions $P_{2, n-1}(s)$ and $\vec{P}_{n-1}(s)$ in Eqs.~\eqref{OmnesSolution1channel} and \eqref{OmnesSolution2channel} are determined by matching to the amplitudes derived from the leading order chiral Lagrangian at low energies when the $\pi\pi$ FSI is switched off~\cite{Chen:2015jgl,Chen:2019mgp}. This introduces four parameters.
For the contribution from triangle diagrams with the $D$-wave $D_1D^*\pi$ coupling, three additional subtraction constants are introduced since the leading order chiral Lagrangian~\cite{Chen:2015jgl,Chen:2019mgp} does not have the required tensor structure.  Details of the Lagrangians~\cite{Wang:2013cya,Guo:2013zbw,Guo:2020oqk,Mehen:2013mva}, the subtractions and the amplitudes are provided in the Supplemental Material.\footnote{See Supplemental Material for further details on the calculations of the amplitudes, and additional fits and fit parameters.}

As for the $J/\psi\pi$-$ D\bar{D}^\ast$ coupled-channel system, we consider the $S$-wave interaction respecting unitarity.
By denoting the $J/\psi\pi$ and $ D\bar{D}^\ast$ channels with 1 and 2,
respectively, the corresponding coupled-channel
$T$-matrix can be obtained as
\begin{equation}
T = (\mathbb{I}-V \cdot G)^{-1} \cdot V~,
\end{equation}
where $G=\text{diag}\{G_1(t),G_2(t)\}$ is the two-point loop function diagonal matrix and the
potential can be written as
$V_{ij} = 4 \sqrt{m_{i1} m_{i2}} \sqrt{m_{j1} m_{j2}}\,  C_{ij}$,
with $m_{i,n}$ the mass of the $n$th particle in
channel $i$.
The $J/\psi \pi \to J/\psi \pi$ interaction is known to have a
tiny scattering length~\cite{Yokokawa:2006td,Liu:2012dv}, and can be neglected by setting $C_{11}=0$.
For the $ D\bar{D}^\ast \to J/\psi \pi$ $S$-wave interaction, we use
the simplest possible assumption by taking it as a constant.
For the $ D\bar{D}^\ast \to  D\bar{D}^\ast$ scattering, we allow an energy-dependent term by setting
$C_{22}=C_{1Z}+b[t-(M_D+M_{D^\ast})^2)]/[2(M_D+M_{D^\ast})]$~\cite{Albaladejo:2015lob,Du:2022jjv}, where $t$ is the square of the total c.m. energy of $D\bar D^* $, and $C_{1Z}$ and $b$ are two parameters.
The energy-dependence term is introduced to allow for the $Z_c(3900)$ pole to be above the $D\bar D^*$ threshold.
The SU(3) light-flavor symmetry is employed to equal the potential of the $J/\psi K$-$D\bar{D}^{\ast}_{s}$ to that of $J/\psi \pi$-$D\bar{D}^*$~\cite{Du:2022jjv}.
We use the dimensionally regularized two-point scalar loop function~\cite{Oller:1998zr},
\begin{eqnarray}\label{eq.Gfunction}
G_i(t)&=&\frac1{16\pi^2}\bigg\{a_i(\mu)+\log\frac{m_{i1}^2}{\mu^2}+\frac{m_{i2}^2-m_{i1}^2+t}{2t} \log\frac{m_{i2}^2}{m_{i1}^2} \nonumber\\
&& +\frac{q_i}{\sqrt{t}}\log\frac{(t+2q_i \sqrt{t})^2-(m_{i1}^2-m_{i2}^2)^2}{(t-2q_i \sqrt{t})^2-(m_{i1}^2-m_{i2}^2)^2}\bigg\},
\end{eqnarray}
where $q_i^2 = \lambda(t,m_{i1}^2,m_{i2}^2)/(4t)$ denotes the c.m. momentum squared for channel $i$, with
the K\"all\'en triangle function $\lambda(a,b,c)=a^2+b^2+c^2-2(ab+ac+bc)$. A variation of the scale $\mu$ can always be absorbed by a corresponding change of the subtraction constants $a_i(\mu)$. The values of $a_1(\mu=1\, \text{GeV}) = -2.77$ and $a_2(\mu=1\, \text{GeV})=-3.0$ are fixed following~\cite{Du:2022jjv}.

For the process $Y(p_a) \to D^\ast(p_b) \bar{D}(p_d)\pi(p_c)$, we consider the Feynman diagrams (2a)-(2d) in Fig.~\ref{fig.FeynmanDiagram}.
The $D^{\ast}\bar{D}$ spectra is then obtained as
\begin{equation}
\frac{\diff\Gamma}{\diff m_{D^*\bar{D}}} = \frac{\mathcal{N} m_{D^{\ast}\bar{D}}}{128\pi^3 M_Y^3}\int_{s_-}^{s_+}
\overline{|\M^{Y \to D^\ast \bar{D}\pi}(s,t)|^2}
\diff s
\,,\label{eq.DstarDmassdistribution}
\end{equation}
where $t=m_{D^*\bar{D}}^2$, $s_\pm$ are the limits of the variable $s=(p_c+p_d)^2$ for the decay,
and $\mathcal{N}$ is an unknown normalization factor. 
The explicit expression for the amplitude of $Y \to D^{\ast}\bar{D} \pi $ is provided in the Supplemental Material.\\

\textbf{3. Results and discussions}\\

With the above constructed amplitudes, which account for TSs from the open-charm triangle diagrams, the $\pi\pi$-$K\bar K$ FSI and the $D\bar D^*_{(s)}$-$J/\psi\pi(K)$ FSI, we fit to the updated BESIII data. The data include both the $\pi^+\pi^-$ and $J/\psi\pi^\pm$ invariant mass distributions of $e^+e^-\to J/\psi\pi^+\pi^-$~\cite{BESIII:2017bua} and the $D^0D^{*-}$ distribution of $e^+e^-\to D^0D^{*-}\pi^+$~\cite{BESIII:2015pqw}.
For $e^+e^-\to J/\psi\pi^+\pi^-$ process, the background events from the experimental analyse~\cite{BESIII:2017bua} are subtracted from the data. While for the $e^+e^-\to D^0D^{*-}\pi^+$ process, there is little background from other processes owing to the use of the double-$D$-tag technique.\footnote{Although the non-$Z_c(3900)$ contribution was denoted as ``background'' and was added incoherently in the BESIII analysis~\cite{BESIII:2015pqw}, it is not from misidentified particles and thus not a genuine background.}
% Since the non-$Z_c(3900)$ events can be described by the $YD^* \bar D\pi $ contact terms and the $D_1$-exchange tree diagram as given in Figs.~\ref{fig.FeynmanDiagram}\,(2a) and (2c), respectively, 
Thus, the BESIII data for $e^+e^-\to D^0D^{*-}\pi^+$~\cite{BESIII:2015pqw} are fitted without any background subtraction. 
As in~\cite{BESIII:2017bua,Pilloni:2016obd}, we consider the measurements at $E=4.23\GeV$ and $4.26\GeV$ (which have 152 and 160 data points, respectively) as independent, and thus the couplings related to the source $Y$ ($y$, $g_{YP1}$, $g_{YP2}$, and $g_{YP3}$) at these two energy points are independent fit parameters. The same holds for the seven subtraction parameters mentioned above.
Both energies, however, share the same $J/\psi\pi$--$D\bar{D}^{\ast}$ coupled-channel $T$-matrix.
In total, there are 27 free parameters: $y$, $g_{YP1}$, $g_{YP2}$, $g_{YP3}$, $\mathcal{N}$ and seven subtraction parameters for each $e^+e^-$ c.m. energy, and common parameters $C_{12}$, $C_{1Z}$ and $b$ for the $J/\psi\pi$-$D\bar{D}^*$ scattering $T$-matrix.
%-------------------
\begin{figure*}[tbh]
  \centering
   \includegraphics[width=\textwidth]{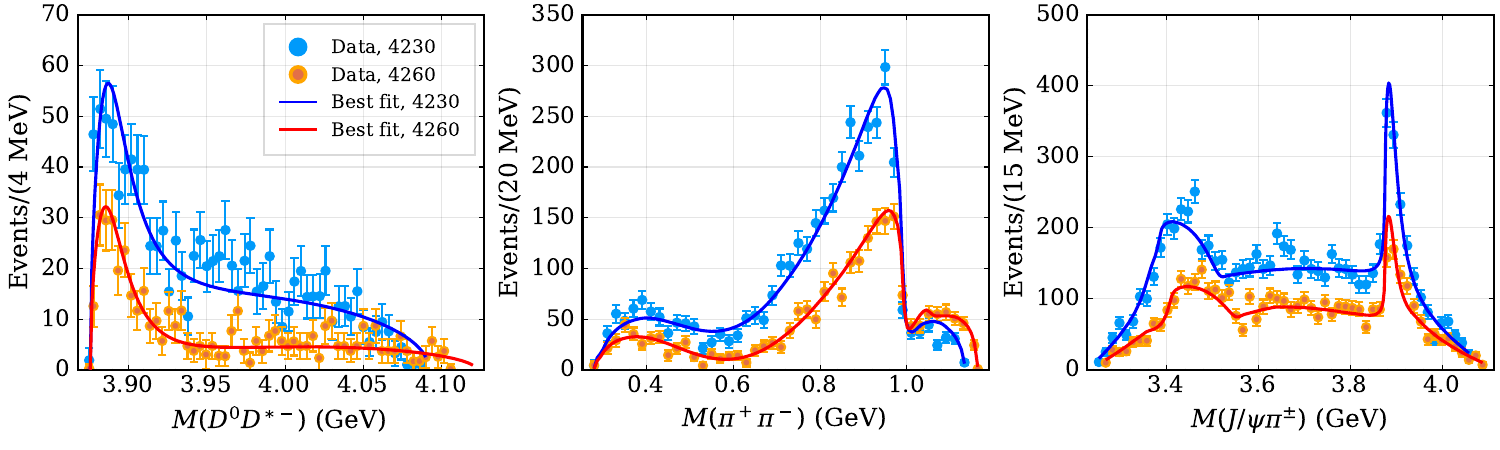}
     \caption{Best fit results in comparison with data of the $D^0 D^{\ast-}$ mass spectra for $e^+e^-
 \rightarrow D^0 D^{\ast-} \pi^+$~\cite{BESIII:2015pqw}, the $\pi^+\pi^-$ and $J/\psi\pi^\pm$ invariant mass spectra for
 $e^+e^- \to J/\psi \pi^+\pi^-$~\cite{BESIII:2017bua} (from left to right) at the $e^+e^-$ c.m. energies $E=4.23$~GeV (blue) and $4.26$~GeV (red). }
   \label{fig.ASandAD}
\end{figure*}
%-------------------
Fits were performed using MINUIT~\cite{James:1975dr,iminuit,iminuit.jl}. 
We do not consider the convolution of the distributions with the energy resolution of the BESIII data since the latter is much more precise than the bin widths of the fitted invariant mass distributions.
The best fit can well describe all the BESIII data (see Fig.~\ref{fig.ASandAD}), with $\chi^2/\text{d.o.f.} =1.66$.
The parameter values and a few other fits with sizeably worse quality are provided in the Supplemental Material.

Since the triangle diagrams have logarithmic singularities near the $D\bar D^*$ threshold, it is a priori unclear whether a $Z_c$ pole is needed to explain the data.
The ($\eta_1\eta_2$) Riemann sheet (RS) of the $J/\psi\pi$-$D\bar{D}^{\ast}$ $T$-matrix can be accessed by analytically continue the $G_i$ function in Eq.~\eqref{eq.Gfunction}:
\begin{align}
G_i(t) & \rightarrow G_i(t)+\eta_i 2 i\rho_i(t)\,\label{eq:RiemannSheetDefinition}
\end{align}
where $\rho_i(t)=\lambda^{1/2}(t,m_{i1}^2,m_{i2}^2)/(16\pi t)$ is the two-body phase space in channel $i~(i=1,2)$. In this convention, the physical sheet is denoted as RS-I=(00), and the other three RSs are labeled as RS-II=(10), RS-III=(11), and RS-IV=(01).

\begin{figure}[tb]
  \centering
  \includegraphics[width=\linewidth]{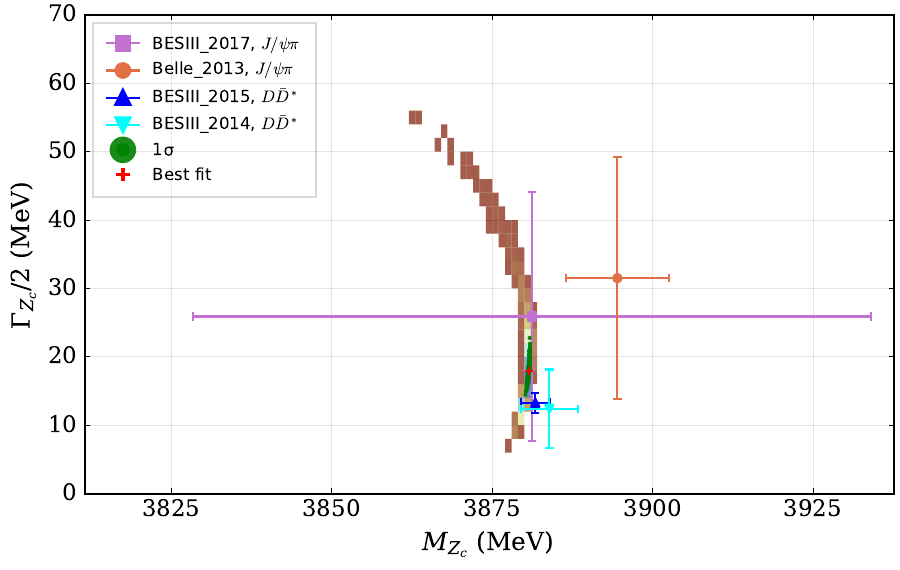}
      \caption{$Z_c$ pole positions. The red cross denotes the best fit result. The green dots are from 320 parameter sets within 1$\sigma$, which are selected from $10^9$ parameter sets sampled randomly following a multivariate normal distribution with the parameter correlations (the brown dots correspond to poles from $3\times10^3$ such parameter sets). The $Z_c(3900)$ mass and width from the BESIII~\cite{BESIII:2017bua,BESIII:2013qmu,BESIII:2015pqw} and Belle~\cite{Liu:2013dau} analyses are shown for comparison.  }
    \label{fig.Zcpole}
\end{figure}

To obtain the statistical uncertainties, we generate $10^9$ parameter sets following a multivariate normal distribution taking into account the parameter correlations. Among them 320 sets are within $1\sigma$, which are selected by requiring the corresponding Mahalanobis distance from the best fit to be $\leq 1$.
We get $C_{12}=-0.006^{+0.006}_{-0.013}$~fm$^2$, $C_{1Z}=(-0.21\pm0.01)$~fm$^2$, and $b=-0.34^{+0.04}_{-0.05}$~fm$^3$.
With these $1\sigma$ parameter sets, there is always a pole on RS-III in the $J/\psi\pi$--$D\bar{D}^{\ast}$ coupled-channel $T$-matrix above the $D\bar{D}^{\ast}$ threshold, which is identified as a resonance corresponding to the $Z_c(3900)^\pm$.
Consequently, the $Z_c$ pole is precisely determined as
\begin{equation}\label{ZcPoleParameters}
  (3880.7 \pm 1.7\pm 22.4) - i (17.9 \pm 0.7\pm 7.7)\text{ MeV}\,.
\end{equation}
Here the first error is statistically propagated from the data, and the corresponding pole distribution is shown as the green dots in Fig.~\ref{fig.Zcpole}, where the mass and half width of $Z_c$ are identified as the real and imaginary parts of the pole.
The second error is systematic and estimated in the following way: We take the list of various sources of systematic errors in the BESIII analysis of the $J/\psi\pi^+\pi^-$ data from Table~II in Ref.~\cite{BESIII:2017bua}, and subtract out those that have been under control by using the dispersive approach, i.e., those involving the light scalar resonances and the nonresonant part for $\pi\pi$, as well as the one corresponding to the $Z_c$ parameterization with a constant Breit-Wigner width. 
These systematic errors are added in quadrature, and the resulting values are taken as the systematic uncertainties.
The so-obtained systematic error for the mass $M_{Z_c}$ is then scaled by the ratio of the systematic errors of the pole mass and $M_{Z_c}$, which is a Flatt\'e parameter, in the BESIII analysis (52.7 and 38.0~MeV, respectively),
and analogously for the width.
Note that the total systematic uncertainties given in the BESIII analysis~\cite{BESIII:2015pqw} for the $e^+e^-
 \rightarrow D^0 D^{\ast-} \pi^+$ reaction are 1.6~MeV and 2.1 MeV for the mass and decay width of $Z_c$, respectively, which are more than one-order-of-magnitude smaller than those in the $e^+e^- \to J/\psi \pi^+\pi^-$ reaction~\cite{BESIII:2017bua}. They are not considered here for a rather conservative estimate.

The parameter $C_{12}$ controlling the $J/\psi\pi$-$D\bar D^*$ channel coupling is not well fixed.
Thus, the fit barely constrains on the ratio of $Z_c$ couplings to the $D\bar D^*$ and $J/\psi \pi$ (derived from the residues of the $T$-matrix elements), which is $|g_{Z_cD\bar D^*}/g_{Z_cJ/\psi\pi}| =118\pm 160_\text{stat}$.

To further investigate the nature of the $Z_c(3900)$ pole, we use the method proposed in~\cite{Matuschek:2020gqe} to calculate the compositeness of the resonances $Z_c(3900)$. The compositeness $\bar X_A$ measuring the probability of finding the two-body component in a near-threshold resonance is given by~\cite{Matuschek:2020gqe}
\begin{equation}
\bar X_A =(1+2|r_0/a_0|)^{-1/2} \,,
\end{equation}
where $a_0$ and $r_0$ are the isoscalar $D\bar D^*$ $S$-wave scattering length and the effective range, respectively.

From the effective range expansion for the $D\bar{D}^*$ scattering amplitude
\begin{equation}\label{eq.EREofTmatrix}
\frac1{T_{22}(k)} = - \frac1{8\pi \sqrt{t}} \left[\frac1{a_0} + \frac12 r_0 k^2 -i\, k + \mathcal{O}(k^4) \right],
\end{equation}
where $k$ refers to the three-momentum of the $\bar{D}^*/D$ in the c.m. frame,
we obtain
\begin{align} \label{eq.compositeness}
a_0 &=\big(1.11\pm 0.16 -i(0.01\pm 0.01)\big) \text{ fm}\,, \nonumber\\
r_0 &=\big(-1.88\pm 0.36+i(0.01\pm0.01)\big) \text{ fm}\,,\nonumber\\
\bar X_A&=0.48\pm 0.02\,,
\end{align}
where the errors are statistical only.
The $\bar X_A$ value is larger than that obtained in~\cite{Yan:2023bwt}, which does not consider the triangle diagrams or the $\pi\pi$ FSI explicitly. Since the former provides some strength to the peaks, its absence could influence the corresponding $D\bar D^*$ interaction strength. The value of $\bar X_A$ being about 0.5 indicates that the $D\bar{D}^\ast$ molecular and additional components are likely to be of similar importance in the $Z_c(3900)$ state. The additional component is of short-distance nature compared to the $D\bar D^*$, whose origin, however, cannot be deduced from the analysis here.\\

\textbf{4. Summary} \\

In this Letter, we provide a simultaneous description of
the $\pi^+\pi^-$ and
$J/\psi\pi^\pm$ invariant mass distributions for $e^+e^-
\rightarrow J/\psi \pi^+\pi^-$ and the $D^0 D^{\ast-}$ mass spectrum for $e^+e^-
\rightarrow D^0 D^{\ast-} \pi^+$ at two energy points $E=4.23$~GeV and $4.26$~GeV.
The open-charm triangle diagrams that contain TS contributions near the physical region, the $J/\psi\pi$-$D\bar D^*$ $S$-wave interactions that are flexible enough to have a $Z_c$ resonance pole, and the strong $\pi\pi$-$K\bar K$ FSI in $e^+e^-\to  J/\psi \pi^+\pi^-$ are all taken into account, with the latter treated using a dispersive approach.
Our best fit leads to a robust and precise extraction of the $Z_c(3900)$ pole position. The pole mass and width are $(3880.7 \pm 1.7_\text{stat}\pm 22.4_\text{syst})$~MeV, and $(35.9 \pm 1.4_\text{stat}\pm 15.3_\text{syst})$~MeV, respectively. The compositeness defined in~\cite{Matuschek:2020gqe} for $Z_c(3900)$ implies that the $D\bar{D}^\ast $ molecular and additional components are of similar importance in the formation of the $Z_c(3900)$ state.
Exploration of possible origins of the short-distance component would shed light on understanding this and other exotic hadrons.

\medskip
\medskip
\medskip

Conflict of interest:

The authors declare that they have no conflict of interest.

\begin{acknowledgements}

We are grateful to Zhi-Qing Liu and Rong-Gang Ping for providing us the BESIII data, and Chang-Zheng Yuan and Zhi-Hui Guo for helpful discussions. This work is supported in part by the National Natural Science Foundation of China (NSFC) under Grants No.~11975028, No.~11974043, No.~12361141819, No. 12125507, No. 11835015, and No. 12047503;
by the Chinese Academy of Sciences under Grant No. YSBR-101 and No. XDB34030000; and by the NSFC and the Deutsche Forschungsgemeinschaft (DFG) through the funds provided to the Sino-German Collaborative Research Center TRR110 ``Symmetries and the Emergence of Structure in QCD'' (NSFC Grant No. 12070131001, DFG Project-ID 196253076).

\end{acknowledgements}

\medskip
\medskip
\medskip

\bibliography{refs}

\begin{onecolumngrid}

\newpage

\begin{center}
% DO NOT EDIT HERE. Instead edit macro in main.tex to keep
  \textbf{\large\normalfont\bfseries\boldmath Precise determination of the pole position of the exotic $Z_c(3900)$} \\
\vspace{0.05in}
{ \it \large Supplementary Material}\\
\vspace{0.05in}
\end{center}

\section{Amplitudes for $ Y \to J/\psi\pi^+\pi^-$}
\label{supp:YtoJpsipipi}

\subsection{Loop amplitudes for $ Y \to J/\psi \pi\pi $ and $J/\psi K\bar{K}$}
\label{supp:loop}

In order to calculate the open-charm loop diagrams,
we need the Lagrangian for the coupling of the $Y$ to $\bar{D}D_1$/$\bar{D}_sD_{s1}$~\cite{Wang:2013cya,Guo:2013zbw},
\begin{align}
\L_{YD_1D}&=\frac{y}{\sqrt{2}}Y^i\left(\bar{D}_{a}^\dag
D_{1a}^{i\dag}-\bar{D}_{1a}^{i\dag} D_a^\dag
\right)+{\rm H.c.} \, ,
\label{LagrangianYD1D}
\end{align}
and that for the coupling of the $D_1$ to $D^\ast\pi$/$D_s^\ast K$, and $D_{s1}$ to $D^\ast K$~\cite{Guo:2020oqk},
\begin{align}
  \L_{D_1D^\ast \phi} = i\frac{h_S}{\sqrt{6}F_\pi} \vec D_{1b} \cdot \vec D^{*\dag}_a
  \partial^0 \Phi_{ba}
 +\frac{h_D}{2F_\pi} {\rm Tr} \left[T_{1b}^i\sigma^j H_a^\dag
 \right] \partial^i\partial^j\Phi_{ba}\,,
\end{align}
where $F_\pi=92.1\MeV$ the pion decay constant, $\Phi$ is the $3\times 3$ matrix containing the pseudo-Nambu-Goldstone
boson octet
\begin{equation}
  \Phi=\left(\begin{array}{ccc}
  \frac{1}{\sqrt{2}} \pi^0+\frac{1}{\sqrt{6}} \eta & \pi^{+} & K^{+} \\
  \pi^{-} & -\frac{1}{\sqrt{2}} \pi^0+\frac{1}{\sqrt{6}} \eta & K^0 \\
  K^{-} & \bar{K}^0 & -\frac{2}{\sqrt{6}} \eta
  \end{array}\right),
\end{equation}
$H_a=\vec{D}_a \cdot \boldsymbol{\sigma}+D_a$,
and $ T_{1a}^i = \sqrt{2/3}\, D_{1a}^i + i \sqrt{1/6}\, \epsilon_{ijk} D_{1a}^j \sigma^k$. Here $a=1,2,3$ represent the mesons with quark content $[c\bar{u}]$, $[c\bar{d}]$, and $[c\bar{s}]$ in order, and $\boldsymbol{\sigma}$ denotes the Pauli matrices in the spinor space.
The coupling constants of the $S$- and $D$-wave $D_1 D^\ast\pi$ vertex are obtained as $|h_S|=0.57$ and $|h_D|=1.17$ GeV$^{-1}$, respectively~\cite{Guo:2020oqk}.

For the contact $ Y D^\ast \bar{D}\pi$ vertex, e.g. in Fig.~\ref{fig.FeynmanDiagram}\,(1c), the Lagrangian at leading order in heavy-meson chiral perturbation theory
reads~\cite{Mehen:2013mva}
\begin{equation}\label{LagrangianJpsiDstarDpi}
\L_{Y D^\ast D \phi}= -\frac{ g_{Y P1}}{\sqrt2 F_\pi } \langle Y\bar{H}_a^\dagger H_b^\dagger  \rangle \partial^0 \Phi \,,
\end{equation}
where
$\bar{H}_a=- \bar{\vec{D}}_a \cdot \boldsymbol{\sigma}+\bar{D}_a$.

The amplitude of the triangle diagram Fig.~(1b) for the $ Y(p_a) \to J/\psi(p_b) \pi(p_c)\pi(p_d)$ processes is
\begin{align}\label{eq.MTriForJpsipipi}
\hat{\M}^{\text{tri},\pi\pi}(s,\theta)=&\,
\frac{y h_S}{\sqrt{6} F_0}\sqrt{M_Y M_D M_{D^\ast}}M_{D_1}\bigg[ p_d^0 I_{D_1DD^\ast}(u)T_{12}(u)+p_c^0 I_{D_1DD^\ast}(t)T_{12}(t) \bigg]  \epsilon_Y \cdot\epsilon_\psi^\ast \notag \\
&\, +\frac{y h_D}{\sqrt{6} F_0}\sqrt{M_Y M_D M_{D^\ast}}M_{D_1} \bigg[I_{D_1DD^\ast}(u)T_{12}(u)(3\vec{p}_d\cdot \epsilon_Y \vec{p}_d\cdot \epsilon_\psi^\ast-\vec{p}_d^2 \epsilon_Y \cdot\epsilon_\psi^\ast) \notag \\
&\,+I_{D_1DD^\ast}(t)T_{12}(t)(3\vec{p}_c\cdot \epsilon_Y \vec{p}_c\cdot \epsilon_\psi^\ast-\vec{p}_c^2 \epsilon_Y \cdot\epsilon_\psi^\ast)\bigg]
\,,
\end{align}
where $\epsilon_{Y}$ and $\epsilon_\psi$ denote the polarization of the $Y$ and $J/\psi$ respectively, $\vec{p}_{c/d}$ is the three-momentum of the $\pi(p_{c/d})$, and $I_{P_1 P_2 P_3}(t)$ is the scalar 3-point loop with intermediate particles $P_1$, $P_2$, and $P_3$. Here the Mandelstam variables $t$ and $u$ are related to $\theta$ by
\begin{align}
  \cos\theta = (t-u)/\kappa(s), \quad \kappa(s)=\sqrt{1-4m_\pi^2/s}\lambda^{1/2}(M_Y^2,s,M_\psi^2).
\end{align}
Evaluated in the rest frame of the initial state $[p=(M_Y,\vec{0})]$, and with all intermediate particles treated nonrelativistically, the scalar three-point loop function in the triangle diagram reaction $A(p) \rightarrow P_1(l)P_2(p-l)\rightarrow B(k)P_3(l-k)P_2(p-l) \rightarrow B(k)C(p-k)$ reads~\cite{Guo:2010ak,Guo:2019twa}
\begin{align}
I_{P_1 P_2 P_3}(t)&\equiv i\int \!\frac{d^4l}{(2\pi)^4}\frac{1}{[l^2-m_1^2+i\epsilon][(p-l)^2-m_2^2+i\epsilon][(l-k)^2-m_3^2+i\epsilon]}\nonumber\\
&\simeq\frac{\mu_{12}\mu_{23}}{16\pi m_1m_2m_3}\frac{1}{\sqrt{a}} \left[\text{arctan}\bigg(\frac{c_2-c_1}{2\sqrt{a(c_1-i\epsilon)}}\bigg)
-\text{arctan}\bigg(\frac{c_2-c_1-2a}{2\sqrt{a(c_2-a-i\epsilon)}}\bigg)\right],
\end{align}
where $t=(p-k)^2$, $\mu_{ij}=(m_im_j)/(m_i+m_j)$, $a=(\mu_{23}q_B/m_3)^2$, $c_1=2\mu_{12}b_{12}$, $c_2=2\mu_{23}b_{23}+q_B^2\mu_{23}/m_3$, with $b_{12}=m_1+m_2-M_Y$, $b_{23}=m_2+m_3+E_B-M_Y$, and $q_B(E_B)$ is the momentum (energy) of the particle $B$ in the $e^+e^-$ c.m. frame.

The bubble amplitude, i.e., Fig.~(1c), for the $ Y(p_a) \to J/\psi(p_b) \pi(p_c)\pi(p_d)$ processes reads
\begin{align}\label{eq.MBubForJpsipipi}
\hat{\M}^{\text{bub},\pi\pi}(s,\theta)=
\frac{2g_{PY1}}{ F_0}\sqrt{M_Y M_D M_{D^\ast}}\bigg[ p_d^0 G_2(u)T_{12}(u)+p_c^0 G_2(t)T_{12}(t) \bigg]  \epsilon_Y \cdot\epsilon_\psi^\ast
\,.
\end{align}

The triangle amplitude for the $ Y(p_a) \to J/\psi(p_b) K(p_c)\bar{K}(p_d)$ processes reads
\begin{align}\label{eq.MTriForJpsiKK}
\hat{\M}^{\text{tri},K\bar{K}}(s,\theta)=&\,
\frac{y h_S}{2\sqrt{6} F_0}p_d^0 T_{12}(u)\bigg[ \sqrt{M_Y M_D M_{D_s^\ast}}M_{D_1} I_{D_1 D D_s^\ast}(u)+\sqrt{M_Y M_{D_s} M_{D^\ast}}M_{D_{s1}} I_{D_{s1} D_s D^\ast}(u) \bigg]  \epsilon_Y \cdot\epsilon_\psi^\ast   \nonumber\\&
+\frac{y h_S}{2\sqrt{6} F_0}p_d^0 T_{12}(t)\bigg[ \sqrt{M_Y M_D M_{D_s^\ast}}M_{D_1} I_{D_1 D D_s^\ast}(t)+\sqrt{M_Y M_{D_s} M_{D^\ast}}M_{D_{s1}} I_{D_{s1} D_s D^\ast}(t) \bigg]  \epsilon_Y \cdot\epsilon_\psi^\ast   \nonumber\\&
+\frac{y h_D}{2\sqrt{6} F_0}T_{12}(u)\bigg[ \sqrt{M_Y M_D M_{D_s^\ast}}M_{D_1} I_{D_1 D D_s^\ast}(u)+\sqrt{M_Y M_{D_s} M_{D^\ast}}M_{D_{s1}} I_{D_{s1} D_s D^\ast}(u) \bigg]
\nonumber\\& \times(3\vec{p}_d\cdot \epsilon_Y \vec{p}_d\cdot \epsilon_\psi^\ast-\vec{p}_d^2 \epsilon_Y \cdot\epsilon_\psi^\ast)
\nonumber\\& +\frac{y h_D}{2\sqrt{6} F_0}T_{12}(t)\bigg[ \sqrt{M_Y M_D M_{D_s^\ast}}M_{D_1} I_{D_1 D D_s^\ast}(t)+\sqrt{M_Y M_{D_s} M_{D^\ast}}M_{D_{s1}} I_{D_{s1} D_s D^\ast}(t) \bigg]
\nonumber\\& \times(3\vec{p}_c\cdot \epsilon_Y \vec{p}_c\cdot \epsilon_\psi^\ast-\vec{p}_c^2 \epsilon_Y \cdot\epsilon_\psi^\ast)
\,.
\end{align}

\subsection{Helicity amplitude for the $Y \to J/\psi \pi\pi$ process \label{sec.helicity} }
\label{supp:helicity}

Formally, the loops amplitude of $Y(p_a) \rightarrow J/\psi(p_d)\pi(p_c)\pi(p_d)$ given in Eqs.~\eqref{eq.MTriForJpsipipi} and~\eqref{eq.MBubForJpsipipi} can be written as
\begin{align}\label{eq.Mhat}
\hat{\M}(s,\theta)&= \hat{M}_A \epsilon_Y \cdot\epsilon_\psi^\ast
+\hat{M}_B^u(3\vec{p}_d\cdot \epsilon_Y \vec{p}_d\cdot \epsilon_\psi^\ast-\vec{p}_d^2 \epsilon_Y \cdot\epsilon_\psi^\ast)
+\hat{M}_B^t(3\vec{p}_c\cdot \epsilon_Y \vec{p}_c\cdot \epsilon_\psi^\ast-\vec{p}_c^2 \epsilon_Y \cdot\epsilon_\psi^\ast)\nonumber\\
&\equiv \hat{\M}_{A}(s,\theta)+\hat{\M}_{B}(s,\theta)
\,.
\end{align}
Here we divide the $\hat{\M}(s,\theta)$ into two parts based on the different tensor structures. The part $\hat{\M}_{A}(s,\theta)$ is purely proportional to $\epsilon_Y \cdot\epsilon_\psi^\ast$, and it is contributed by the sum of the
triangle diagrams with $S$-wave $D_1 D^\ast\pi$ coupling and the bubble diagrams. The other part $\hat{\M}_{B}(s,\theta)$ is contributed by the
triangle diagrams with $D$-wave $D_1 D^\ast\pi$ coupling. The expressions for the $\hat{M}_A$ and $\hat{M}_B^{t(u)}$ can be obtained by comparing Eqs.~\eqref{eq.MTriForJpsipipi}, \eqref{eq.MBubForJpsipipi} and \eqref{eq.MTriForJpsiKK} with Eq.~\eqref{eq.Mhat} straightforwardly. Correspondingly, the rhc part of the amplitude should also contain these two types of tensor structures. The full amplitude
can be written as
\begin{align}\label{eq.MHelicity}
\M^\text{full}(s,\theta)=&\, (M_A+\hat{M}_A) \epsilon_Y \cdot\epsilon_\psi^\ast
+(M_B^u+\hat{M}_B^u)(3\vec{p_d}\cdot \epsilon_Y \vec{p_d}\cdot \epsilon_\psi^\ast-\vec{p_d^2} \epsilon_Y \cdot\epsilon_\psi^\ast) \nonumber\\&
+(M_B^t+\hat{M}_B^t)(3\vec{p_c}\cdot \epsilon_Y \vec{p_c}\cdot \epsilon_\psi^\ast-\vec{p_c^2} \epsilon_Y \cdot\epsilon_\psi^\ast) \nonumber\\
=&\, (H_{\mu\nu}^{\lambda,\lambda^\prime}+\hat{H}_{\mu\nu}^{\lambda,\lambda^\prime})\epsilon_{\lambda}^\mu \epsilon_{\lambda^\prime}^{\ast\nu}\,,
\end{align}
where we reexpress the full amplitude as
helicity amplitudes to facilitate the partial-wave decomposition of the two-pion system. Note that we need not to give the explicit expressions of the rhc part $M_A$ and $M_B^{t(u)}$ in Eq.~\eqref{eq.MHelicity} since only the helicity amplitudes appear in the numerical calculation. Here we focus on the non-relativistic case, and therefore $\mu, \nu=1,2,3$ (in the following, we will use the notations $i,j$) and $\lambda, \lambda^{\prime}=\pm 1,0$.
The appropriate helicity amplitudes $H^{\lambda,\lambda^\prime}$ and $\hat{H}^{\lambda,\lambda^\prime}$ are obtained by inserting explicit expressions for
the polarization vectors $\epsilon_{\lambda^{(\prime)}}$ occurring in the full amplitudes:
\begin{eqnarray}
\epsilon_\pm = \left(
                           \begin{array}{ccc}
                            \mp\frac{1 }{\sqrt{2}}, \frac{- i}{\sqrt{2}}, 0 \\
                           \end{array}
                         \right),\quad
\epsilon_0 = \left(
                           \begin{array}{ccc}
                             0 , 0, 1
                             \end{array}
                         \right).
\end{eqnarray}

Given in the order of $(+,-, 0)$, the rhc and lhc helicity amplitudes in matrix form read
\begin{align}\label{eq.Hmatrix}
H^{\lambda,\lambda^\prime} &=\left(\begin{array}{ccc}
H_A^{++}+H_B^{++} & H_B^{+-} & H_B^{+0} \\
H_B^{-+} & H_A^{--}+H_B^{--} & H_B^{-0} \\
H_B^{0+} & H_B^{0-} & H_A^{00}+H_B^{00}
\end{array}\right),\notag \\
\hat{H}^{\lambda,\lambda^\prime} &=\left(\begin{array}{ccc}
\hat{H}_A^{++}+\hat{H}_B^{++} & \hat{H}_B^{+-} & \hat{H}_B^{+0} \\
\hat{H}_B^{-+} & \hat{H}_A^{--}+\hat{H}_B^{--} & \hat{H}_B^{-0} \\
\hat{H}_B^{0+} & \hat{H}_B^{0-} & \hat{H}_A^{00}+\hat{H}_B^{00}
\end{array}\right),
\end{align}
respectively, with
\begin{align}
H_A^{++}=&\,H_A^{--}=H_A^{00}=M_A, \\
 H_B^{++}=&\,H_B^{--} =-\frac{1}{2} H_B^{00} =-\frac{1}{8|\mathbf{p}|_d^2}\left(M_B^t\left[2\left(|\mathbf{p}|_d^2+|\mathbf{p}|_c^2-|\mathbf{p}|_d^2\right)^2+\lambda\left(|\mathbf{p}|_d^2,|\mathbf{p}|_c^2,|\mathbf{p}|_d^2\right)\right]\right. \nonumber\\& \left.+M_B^u\left[2\left(|\mathbf{p}|_d^2-|\mathbf{p}|_c^2+|\mathbf{p}|_d^2\right)^2+\lambda\left(|\mathbf{p}|_d^2,|\mathbf{p}|_c^2,|\mathbf{p}|_d^2\right)\right]\right), \\
 H_B^{+-}=&\,H_B^{-+}=-\frac{3 (M_B^t+M_B^u)}{8|\mathbf{p}|_d^2} \lambda\left(|\mathbf{p}|_d^2,|\mathbf{p}|_c^2,|\mathbf{p}|_d^2\right), \\
 H_B^{+0}=&-H_B^{0+} =H_B^{-0}=-H_B^{0-}=-\frac{3 i }{4 \sqrt{2}|\mathbf{p}|_d^2} \sqrt{-\lambda\left(|\mathbf{p}|_d^2,|\mathbf{p}|_c^2,|\mathbf{p}|_d^2\right)}\left(M_B^t\left(|\mathbf{p}|_d^2+|\mathbf{p}|_c^2-|\mathbf{p}|_d^2\right)\right.  \nonumber\\
& \left.-M_B^u\left(|\mathbf{p}|_d^2-|\mathbf{p}|_c^2+|\mathbf{p}|_d^2\right)\right)  ,
\end{align}
and
\begin{align}
\hat{H}_A^{++}=&\,\hat{H}_A^{--}=\hat{H}_A^{00}=\hat{M}_A, \\
 \hat{H}_B^{++}=&\,\hat{H}_B^{--} =-\frac{1}{2} \hat{H}_B^{00}
=-\frac{1}{8|\mathbf{p}|_d^2}\left(\hat{M}_B^t\left[2\left(|\mathbf{p}|_d^2+|\mathbf{p}|_c^2-|\mathbf{p}|_d^2\right)^2+\lambda\left(|\mathbf{p}|_d^2,|\mathbf{p}|_c^2,|\mathbf{p}|_d^2\right)\right]\right. \nonumber\\& \left.+\hat{M}_B^u\left[2\left(|\mathbf{p}|_d^2-|\mathbf{p}|_c^2+|\mathbf{p}|_d^2\right)^2+\lambda\left(|\mathbf{p}|_d^2,|\mathbf{p}|_c^2,|\mathbf{p}|_d^2\right)\right]\right), \\
 \hat{H}_B^{+-}=&\,\hat{H}_B^{-+}=-\frac{3 (\hat{M}_B^t+\hat{M}_B^u)}{8|\mathbf{p}|_d^2} \lambda\left(|\mathbf{p}|_d^2,|\mathbf{p}|_c^2,|\mathbf{p}|_d^2\right), \\
 \hat{H}_B^{+0}=&-\hat{H}_B^{0+} =\hat{H}_B^{-0}=-\hat{H}_B^{0-}=-\frac{3 i }{4 \sqrt{2}|\mathbf{p}|_d^2} \sqrt{-\lambda\left(|\mathbf{p}|_d^2,|\mathbf{p}|_c^2,|\mathbf{p}|_d^2\right)}\left(\hat{M}_B^t\left(|\mathbf{p}|_d^2+|\mathbf{p}|_c^2-|\mathbf{p}|_d^2\right)\right.  \nonumber\\
& \left.-\hat{M}_B^u\left(|\mathbf{p}|_d^2-|\mathbf{p}|_c^2+|\mathbf{p}|_d^2\right)\right)  .
\end{align}
One observes that there are eight independent terms: $H_A^{++}, H_B^{++}, H_B^{+-}$, $H_B^{+0}$, $\hat{H}_A^{++}, \hat{H}_B^{++}, \hat{H}_B^{+-}$, and $\hat{H}_B^{+0}$.
In the $H^{\lambda,\lambda^\prime}$ and $\hat{H}^{\lambda,\lambda^\prime}$ matrices given in Eq.~\eqref{eq.Hmatrix}, the $H_A$ and $\hat{H}_A$ terms only appear in the diagonal elements. Using the relations $H_A^{++}=H_A^{--}=H_A^{00}$, $H_B^{00}=-2 H_B^{++}=-2 H_B^{--}$, $\hat{H}_A^{++}=\hat{H}_A^{--}=\hat{H}_A^{00}$, and $\hat{H}_B^{00}=-2 \hat{H}_B^{++}=-2 \hat{H}_B^{--}$, one obtains
\begin{align}
 &\left|H_A^{++}+H_B^{++}+\hat{H}_A^{++}+\hat{H}_B^{++}\right|^2+\left|H_A^{--}+H_B^{--}+\hat{H}_A^{--}+\hat{H}_B^{--}\right|^2+\left|H_A^{00}+H_B^{00}+\hat{H}_A^{00}+\hat{H}_B^{00}\right|^2\nonumber\\
=&\, 2\left|H_A^{++}+H_B^{++}+\hat{H}_A^{++}+\hat{H}_B^{++}\right|^2+\left|H_A^{++}-2 H_B^{++}+\hat{H}_A^{++}-2 \hat{H}_B^{++}\right|^2 \nonumber\\
=&\, 3\left|H_A^{++}+\hat{H}_A^{++}\right|^2+6\left|H_B^{++}+\hat{H}_B^{++}\right|^2\,,
\end{align}
where no interference between $(H_A+\hat{H}_A)$ and $(H_B+\hat{H}_B)$ survives in the squared amplitudes.

Now we briefly discuss the $(H_A^{\lambda,\lambda^\prime}+\hat{H}_A^{\lambda,\lambda^\prime})$ and $(H_B^{\lambda,\lambda^\prime}+\hat{H}_B^{\lambda,\lambda^\prime})$ terms which can be treated independently
in the dispersive approach. Typically, the unitary conditions read as follows
\begin{align}\label{eq.unitarityTypically}
\textrm{Im}\, (H_A^{\lambda,\lambda^\prime}+H_B^{\lambda,\lambda^\prime})= T^\ast\sigma\left[H_A^{\lambda,\lambda^\prime}+H_B^{\lambda,\lambda^\prime} +\hat{H}_A^{\lambda,\lambda^\prime}+\hat{H}_B^{\lambda,\lambda^\prime}  \right]\,,
\end{align}
where $T$ denotes the scattering amplitude, and $\sigma$ is the phase space factor.
Specifically,
\begin{align}\label{eq.unitarity++}
\textrm{Im}\, (H_A^{++}+H_B^{++})&= T^\ast\sigma\left[H_A^{++}+H_B^{++} +\hat{H}_A^{++}+\hat{H}_B^{++}  \right],\\
\textrm{Im}\, (H_A^{00}+H_B^{00})&= T^\ast\sigma\left[H_A^{00}+H_B^{00} +\hat{H}_A^{00}+\hat{H}_B^{00}  \right].\label{eq.unitarity00}
\end{align}
Using the relations $H_A^{++}=H_A^{00}$, $H_B^{00}=-2 H_B^{++}$, $\hat{H}_A^{++}=\hat{H}_A^{00}$, and $\hat{H}_B^{00}=-2 \hat{H}_B^{++}$, Eq.~\eqref{eq.unitarity00} can be rewritten as
\begin{align}
\textrm{Im}\, (H_A^{++}-2H_B^{++})= T^\ast\sigma\left[H_A^{++}-2H_B^{++} +\hat{H}_A^{++}-2\hat{H}_B^{++}  \right].\label{eq.unitarity00new}
\end{align}
Combining Eqs.~\eqref{eq.unitarity++} and~\eqref{eq.unitarity00new}, we obtain
\begin{align}\label{eq.unitarity++newnew}
\textrm{Im}\, H_A^{++}&= T^\ast\sigma\left[H_A^{++}+\hat{H}_A^{++} \right],\\
\textrm{Im}\, H_B^{++}&= T^\ast\sigma\left[H_B^{++}+\hat{H}_B^{++}  \right].
\end{align}
Note that the $(H_A^{\lambda,\lambda^\prime}+\hat{H}_A^{\lambda,\lambda^\prime})$ term does not interfere with the $(H_B^{\lambda,\lambda^\prime}+\hat{H}_B^{\lambda,\lambda^\prime})$ term in the unitary conditions, and therefore they can be treated separately in the dispersive approach. This is a direct consequence of
\bea
\sum_\text{pol.} \left( \epsilon_Y\cdot\epsilon_\psi^*\right)^*\left(3\vec{p}_{c/d}\cdot\epsilon_Y \vec{p}_{c/d}\cdot\epsilon_\psi^*-\vec{p}_{c/d}^2\epsilon_Y\cdot\epsilon_\psi^*\right)=0.
\eea

The partial-wave projection of the $Y \to J/\psi PP$ helicity amplitudes is given as follows
\begin{equation}\label{eq.partialwaveamp}
\hat{H}^{\lambda,\lambda^\prime,l}(s)=\frac{2 l+1}{2} \int \mathrm{d} \cos \theta \, d_{\lambda-\lambda^\prime, 0}^{l}(\theta) \hat{H}^{\lambda,\lambda^\prime}(s,\theta)\,.
\end{equation}

\subsection{Subtraction terms in the dispersion relations }
\label{supp:subtraction}

In order to determine the necessary number of subtractions to
guarantee the dispersive integrals in
Eqs.~\eqref{OmnesSolution1channel} and \eqref{OmnesSolution2channel} convergent, we need to analyze the high-energy behavior of the integrands.
We have checked that in the intermediate energy region of $1\GeV^2 \lesssim s
\lesssim 100\GeV^2$, for the $\pi\pi$ $S$-wave, the integrand $\Omega^{-1}(x)T_0^0(x)\Sigma(x)\vec{\hat{H}}^{\lambda, \lambda^\prime,0}(s)$ in Eq.~\eqref{OmnesSolution2channel}
grows at most linearly in $s$, and thus two subtractions are sufficient
to make the dispersive integrals convergent. For the $\pi\pi$ $D$-wave, the integrand $\hat{H}_\pi^{\lambda, \lambda^\prime,2}(s)\sin\delta_2^0(x) /|\Omega_2^0(x)|$ in Eq.~\eqref{OmnesSolution1channel} $\sim s^0$ at large $s$, and therefore one subtraction is sufficient to guarantee the convergence of the dispersive integral.
On the other hand, as mentioned above the terms with different tensor structures, namely the $(H_A^{\lambda,\lambda^\prime}+\hat{H}_A^{\lambda,\lambda^\prime})$ term and the $(H_B^{\lambda,\lambda^\prime}+\hat{H}_B^{\lambda,\lambda^\prime})$ term, do not interfere with each other in the unitary conditions. Therefore, they can be treated separately in the dispersive approach.

First we consider the $(H_A^{\lambda,\lambda^\prime}+\hat{H}_A^{\lambda,\lambda^\prime})$ term which is purely proportional to $\epsilon_Y \cdot\epsilon_\psi^\ast$.
Since at low energies, the amplitude should agree with the leading chiral results, the subtraction terms can be determined by matching the chiral contact terms.
As the internal structure of the initial state $Y$ is not yet clear, we follow Ref.~\cite{Chen:2019mgp} to split it into a flavor SU(3) singlet part and an octet part, whose matrix form in the SU(3) flavor space may be written as~\cite{Chen:2019mgp}
\begin{equation}
\label{eq.YComponents} |Y\rangle= a |\mathbbm{V}_1\rangle+ \sqrt{1-a^2}|\mathbbm{V}_8\rangle =  \frac{a}{\sqrt{3}} V_1 \cdot \mathbbm{1}+\frac{\sqrt{1-a^2}}{\sqrt{6}} V_8\cdot \text{diag} \left( 1,  1, - 2\right) ,
\end{equation}
where the indices $1$ and $8$ denote the singlet and octet, respectively.
The effective Lagrangian for the contact $Y\to J/\psi\pi\pi$ and
$Y\to J/\psi K\bar{K}$ couplings, at the lowest order in the chiral expansion
and respecting the heavy-quark spin symmetry,
reads~\cite{Chen:2019mgp,Mannel:1995jt,Chen:2015jgl,Chen:2016mjn,Chen:2019gty}
\begin{equation}\label{LagrangianYpsipipi}
\L_{Y\psi\Phi\Phi} = g_1\langle \mathbbm{V}_{1}^\alpha J^\dag_\alpha \rangle \langle u_\mu
u^\mu\rangle +h_1\langle \mathbbm{V}_{1}^{\alpha} J^\dag_\alpha \rangle \langle u_\mu u_\nu\rangle
v^\mu v^\nu +g_8\langle  J^\dag_\alpha \rangle \langle \mathbbm{V}_{8}^{\alpha} u_\mu u^\mu\rangle
+h_8\langle J^\dag_\alpha \rangle \langle \mathbbm{V}_{8}^{\alpha} u_\mu u_\nu\rangle v^\mu v^\nu
+\mathrm{H.c.}\,,
\end{equation}
where $\langle\ldots\rangle$ represents the trace in the SU(3) flavor space, $J= (\psi/\sqrt{3}) \cdot \mathbbm{1}$, and
$v^\mu=(1,\vec{0})$ denotes the velocity of the heavy quark.
By making use of the Lagrangian
Eq.~\eqref{LagrangianYpsipipi}, one obtains the amplitudes as
\begin{align}
\M^{\chi,\pi\pi}(s,\theta)&=M^{\chi,\pi\pi}(s,\theta)\epsilon_Y \cdot\epsilon_\psi^\ast =-\frac{4}{F_0^2}\sqrt{M_{Y}M_{\psi}}\bigg[\Big(g_1+\frac{g_8}{\sqrt{2}}\Big)p_c\cdot
p_d +\Big(h_1+\frac{h_8}{\sqrt{2}}\Big)p_c^0 p_d^0 \bigg]\epsilon_Y \cdot\epsilon_\psi^\ast\,, \notag\\
\label{eq.ContactPi+KRaw}
\M^{\chi,K\bar{K}}(s,\theta)&=M^{\chi,K\bar{K}}(s,\theta)\epsilon_Y \cdot\epsilon_\psi^\ast =-\frac{4}{F_0^2}\sqrt{M_{Y}M_{\psi}}\bigg[\Big(g_1-\frac{g_8}{2\sqrt{2}}\Big)p_c\cdot
p_d +\Big(h_1-\frac{h_8}{2\sqrt{2}}\Big)p_c^0 p_d^0 \bigg]\epsilon_Y \cdot\epsilon_\psi^\ast\,.
\end{align}
Note that the chiral contact terms in
Eq.~\eqref{eq.ContactPi+KRaw} is purely proportional to $\epsilon_Y \cdot\epsilon_\psi^\ast$, and at low energies the amplitudes
$\vec{H}_{A}^{\lambda, \lambda^\prime,0}(s)$ and $H_{A}^{\lambda, \lambda^\prime,2}(s)$ should match to those from chiral amplitudes.
Namely, in the chiral limit,
the subtraction terms should agree with the partial-wave projection of the low-energy chiral amplitudes
given in Eq.~\eqref{eq.ContactPi+KRaw}.
Therefore, for the $\pi\pi$
$S$-wave, the integral equations take the form
\begin{equation}\label{M02channel}
\vec{H}_{A}^{\lambda, \lambda^\prime,0}(s)=\Omega(s)\bigg\{\vec{M}^{\chi,0}(s)+\frac{s^3}{\pi}\int_{4m_\pi^2}^\infty
\frac{\diff
x}{x^3}\frac{\Omega^{-1}(x)T_0^0(x)\Sigma(x)\hat{\vec{H}}_{A}^{\lambda, \lambda^\prime,0}(s)}{x-s}\bigg\}
\,,
\end{equation}
where $ \vec{M}^{\chi}_0(s)=\big(
   M^{\chi,\pi\pi,0}(s),
   2/\sqrt{3}\,M^{\chi,K\bar{K},0}(s)
   \big)^{T}$, while
for the $D$-wave, it can be written as follows
\be\label{M21channel}
H_{A}^{\pi,\lambda, \lambda^\prime,2}(s)=\Omega_2^0(s)\bigg\{M^{\chi,\pi\pi,2}(s)+\frac{s^3}{\pi}\int_{4m_\pi^2}^\infty
\frac{\diff x}{x^3} \frac{\hat{H}_{A}^{\pi,\lambda, \lambda^\prime,2}(s)\sin\delta_2^0(x)}{|\Omega_2^0(x)|(x-s)}\bigg\} \,.
\ee
Note that in Eqs.~\eqref{M02channel} and~\eqref{M21channel}, oversubtracted dispersion relations have been implemented since $M^{\chi,\pi\pi(K\bar{K}),0}(s)$ and $M^{\chi,\pi\pi,2}(s)$ all grow
as $s^2$.

Next we discuss the $\hat{H}_B^{\lambda,\lambda^\prime}$ term that is produced by the triangle diagrams with $D$-wave $D_1 D^\ast\pi$ coupling.
Since there is no corresponding term with the same tensor structure in the chiral contact Lagrangian, which is constructed at the leading order of the heavy quark expansion, here we introduce new subtraction terms by hand. Namely, for the $\pi\pi$
$S$-wave, the integral equations are written as
\begin{align}\label{M02channel.DwaveD1Dstarpi}
\vec{H}_{B}^{\lambda, \lambda^\prime,0}(s)=\Omega(s)\bigg\{\vec{P}^{\lambda, \lambda^\prime}(s)+\frac{s^2}{\pi}\int_{4m_\pi^2}^\infty
\frac{\diff
x}{x^2}\frac{\Omega^{-1}(x)T_0^0(x)\Sigma(x)\hat{\vec{H}}_{B}^{\lambda, \lambda^\prime,0}(s)}{x-s}\bigg\}
\,,
\end{align}
where $ \vec{P}^{\lambda, \lambda^\prime}(s)=\big(j_1^{\lambda, \lambda^\prime}+j_2^{\lambda, \lambda^\prime} s,
   2(j_3^{\lambda, \lambda^\prime}+j_4^{\lambda, \lambda^\prime} s)/\sqrt{3}\big)$.
For the $\pi\pi$ $D$-wave, the integral equation can be written as
\be\label{M21channel.DwaveD1Dstarpi}
H_{B}^{\pi,\lambda, \lambda^\prime,2}(s)=\Omega_2^0(s)\bigg\{j_5^{\lambda, \lambda^\prime}+\frac{s}{\pi}\int_{4m_\pi^2}^\infty
\frac{\diff x}{x} \frac{\hat{H}_{B}^{\pi,\lambda, \lambda^\prime,2}(s)\sin\delta_2^0(x)}{|\Omega_2^0(x)|(x-s)}\bigg\} \,.
\ee
As stated above, there are three independent $\hat{H}_B^{\lambda,\lambda^\prime}$ terms: $\hat{H}_B^{++}, \hat{H}_B^{+-}$, and $\hat{H}_B^{+0}$. To reduce the number of free parameters in the fits, here we only take into account the subtraction term corresponding to $\hat{H}_B^{++}$, but set the subtraction terms corresponding to $\hat{H}_B^{+-}$  and $\hat{H}_B^{+0}$ to be zero. Namely we choose $j_i=j_i^{++} (i=1,...,5)$.
In the actual fits, we find that strong correlations exist between $j_1$ and $j_2$, $j_3$ and $j_4$, respectively. Therefore we can set $j_2=j_4=0$ in the fits, and we have checked that this setting has negligible effect on the value of $\chi^2$ of the obtained best fits.

\section{Amplitude for $ Y \to D^\ast \bar{D}\pi$}
\label{supp:YtoDstarDpi}

As shown in Fig.~\ref{fig.ASandAD}, the tail of the $D\bar D^*$ spectra data in the $E=4230$~MeV case is much higher than the tail in the $E=4260$~MeV case, therefore we introduce two additional $YD^* \bar D\pi $ contact terms in order to describe the energy dependence of the $D\bar D^*$ spectra, 
\begin{equation}\label{LagrangianJpsiDstarDpinew}
\L_{Y D^\ast D \phi}^\prime= -i\frac{ g_{Y P2}}{\sqrt2 F_\pi } \langle Y (\partial_i \partial_i\bar{H}_a^\dagger) H_b^\dagger  \rangle  \Phi 
-i\frac{ g_{Y P3}}{\sqrt2 F_\pi } \langle Y \bar{H}_a^\dagger(\partial_i \partial_i H_b^\dagger)  \rangle  \Phi \,.
\end{equation}
% Note that in the $e^+e^-\rightarrow J/\psi \pi^+\pi^-$ process, since the effects of the triangle singularity and the $J/\psi\pi$-$D\bar D^*$ coupled-channel interaction have already been taken into account, we do not need to consider the effect of above two additional $YD^* \bar D\pi $ vertexes.

The decay amplitude for the $ Y(p_a) \to D^\ast(p_b) \bar{D}(p_d)\pi(p_c)$ processes read
\begin{align}\label{eq.MForDstarDpi}
\M^{Y \to D^\ast \bar{D}\pi}(s,t,u)=&\,
\frac{y h_S \sqrt{M_Y M_D M_{D^\ast}}M_{D_1}}{2\sqrt{3} F_0}p_c^0
\bigg[ - \frac{1}{u-M_{D_1}^2}+I(t)T_{22}(t) \bigg]  \epsilon_Y \cdot\epsilon_{D^\ast}^\ast   \nonumber\\&
+\frac{y h_D \sqrt{M_Y M_D M_{D^\ast}}M_{D_1}}{2\sqrt{3} F_0}
\bigg[ - \frac{1}{u-M_{D_1}^2}+I(t)T_{22}(t) \bigg]   (3\vec{p}_c\cdot \epsilon_Y \vec{p}_c\cdot \epsilon_{D^\ast}^\ast-\vec{p}_c^2 \epsilon_Y \cdot\epsilon_{D^\ast}^\ast)
\nonumber\\& +
\frac{\sqrt{2}  \sqrt{M_Y M_D M_{D^\ast}}}{F_0}
(g_{PY1}p_c^0-g_{PY2}\vec{p}_d^2-g_{PY3}\vec{p}_c^2)\bigg[ -1+G_{2}(t)T_{22}(t) \bigg]  \epsilon_Y \cdot\epsilon_{D^\ast}^\ast\,.
\end{align}

\section{Additional fits and fit parameters}
\label{supp:additionalfits}

In order to illustrate the effect of the energy-dependence of the $D \bar{D}^\ast  \to D \bar{D}^\ast $ potential and the effect of the $D$-wave $D_1 D^\ast\pi$ coupling, we perform three additional fits. Specifically,
in Fit II we consider both the $S$- and $D$-wave $D_1D^*\pi$ couplings with energy-independent $\bar{D}^\ast D$ interaction ($b=0$); in Fits III and IV, we only consider the $S$-wave $D_1 D^\ast\pi$ coupling with energy-dependent and energy-independent $\bar{D}^\ast D$ interaction, respectively. We denote the best fit in the main text as Fit I, which considers both the $S$- and $D$-wave $D_1D^*\pi$ couplings with energy-dependent $\bar{D}^\ast D$ interaction.

Fit I has $\chi^2/\text{d.o.f.} = 1.66$.
The fit results of Fits II, III, and IV (with $\chi^2/\text{d.o.f.} = 2.18, 1.98$ and $2.51$, respectively, sizeably larger than that of Fit I) are shown in Fig.~\ref{fig.AS}.
The parameter values from all fits are listed in Table~\ref{tablepar1}. For Fit~I, as mentioned in the main text, we randomly sample $1.0\times10^8$ parameter sets following a multivariate normal distribution with the parameter correlations taken into account, and the parameter uncertainties are obtained by selecting only the parameter sets leading to a Mahalanobis distance from the best fit to be $\leq 1$. There are only 406 such parameter sets.
The statistical correlations of the $J/\psi\pi$-$D\bar D^*$ $T$-matrix parameters are given in Table~\ref{tablepar_T}.
For Fits~II, III, and IV, since they are not the preferred ones, we simply take the uncertainties obtained using the MIGRAD algorithm of MINUIT~\cite{James:1975dr,iminuit,iminuit.jl}.

First, one observes that in Fits III and IV only considering the $S$-wave $D_1 D^\ast\pi$ coupling, the $\pi\pi$ spectra of the $E=4.23$~GeV data set in the region above about 0.45~GeV cannot be well described. This indicates the necessity of taking into account the $D$-wave part of the $D_1D^*\pi$ coupling. In contrast, in Fits I and II, including the $D$-wave $D_1D^*\pi$ coupling, the fit qualities are improved significantly.

Second, for Fit II with energy-independent $D\bar{D}^\ast$ potential, we also study the pole structure of the $J/\psi\pi$--$D\bar{D}^{\ast}$ $T$-matrix. Using the central value of the parameters, we find a pole located at 3872.7~MeV (with a tiny imaginary part) below the $ D\bar{D}^\ast$ threshold on RS-IV, which would be a $D\bar{D}^\ast $ virtual state if the $J/\psi\pi$ channel is switched off.

\begin{figure*}[!htb]
 \centering
  \includegraphics[width=\textwidth]{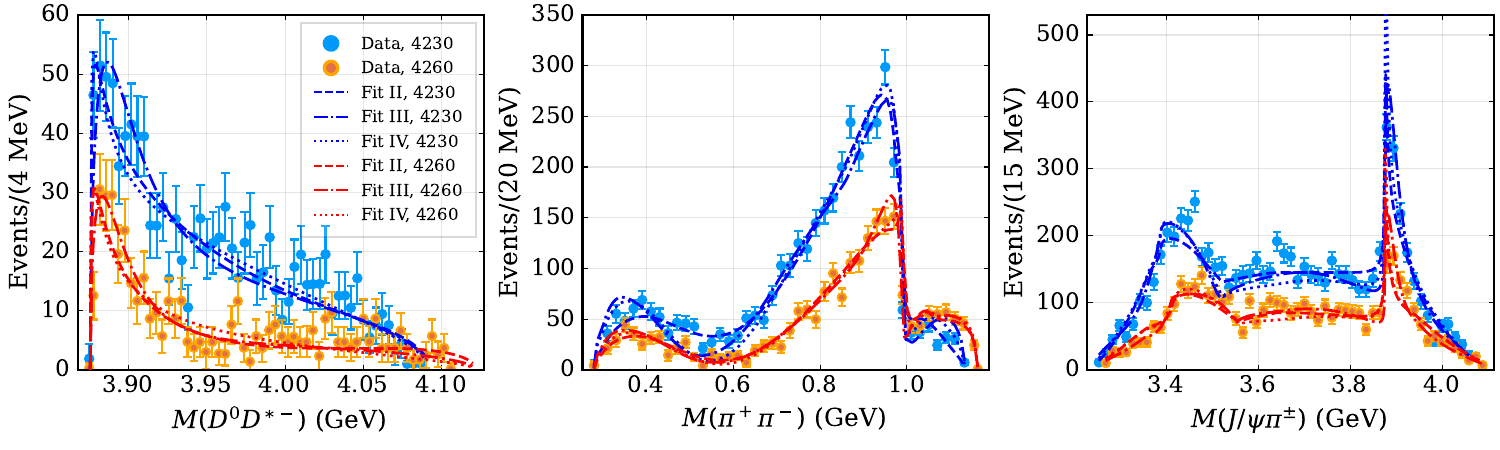}
    \caption{Additional fit results of the $D^0 D^{\ast-}$ mass spectra for $e^+e^-
\rightarrow D^0 D^{\ast-} \pi^+$, the $\pi^+\pi^-$ and $J/\psi\pi^\pm$ invariant mass spectra for
$e^+e^- \to J/\psi \pi^+\pi^-$ (from left to right) at the $e^+e^-$ c.m. energies $E=4.23$ GeV and $E=4.26$ GeV. The results of Fits II, III, and IV are shown as the dashed, dot-dashed, and dotted lines, respectively. The data of the $D^0 D^{\ast-} \pi^+$ final state and the $J/\psi \pi^+\pi^-$ final state are taken from Refs.~\cite{BESIII:2015pqw} and~\cite{BESIII:2017bua}, respectively.  }
  \label{fig.AS}
\end{figure*}

\begin{table}[tbh]
\caption{\label{tablepar1} Fit parameters from the simultaneous fit of the $D^0 D^{\ast-}$ mass spectra for the $e^+e^-
\rightarrow D^0 D^{\ast-} \pi^+$ process and the $\pi^+\pi^-$ and
$J/\psi\pi^\pm$ invariant mass distributions for the $e^+e^-
\rightarrow J/\psi \pi^+\pi^-$ process at $E=4.23$ GeV and $E=4.26$ GeV. 
The corresponding energy-specific parameters are labeled by upper indices $^{(4230)}$ and $^{(4260)}$ for the $E=4.23$ GeV and $E=4.26$ GeV data sets, respectively.
Fit I: considering both the $S$- and $D$-wave $D_1D^*\pi$ couplings, $b\neq0$; Fit II: considering both the $S$- and $D$-wave $D_1D^*\pi$ couplings, $b=0$; Fit III: considering only the $S$-wave $D_1D^*\pi$ coupling, $b\neq0$;
 Fit IV: considering only the $S$-wave $D_1D^*\pi$ coupling, $b=0$. Fit I has a $\chi^2$ value much smaller than all the other fits and is considered to be the main result reported in the text. The parameter errors in Fit I are obtained from a random sampling analysis (see above). The uncertainties of the parameters in the other three fits are obtained using the MIGRAD algorithm in MINUIT.}
\begin{ruledtabular}
\renewcommand{\arraystretch}{1.35}
\begin{tabular}{l|cccc}
        & Fit I & Fit II & Fit III & Fit IV  \\
\hline
$g_1^{(4230)}\times 10^{3}~[\text{GeV}^{-1}]$     &   $ 12.6^{+0.8}_{-0.9}$&   $ 19.3\pm 0.1$&   $ 3.0\pm 0.8$&  $ 10.7\pm 0.2$ \\
$g_8^{(4230)}/g_1^{(4230)}$    &   $ -0.352^{+0.080}_{-0.096}$&$ -0.452\pm 0.015$ & $ 3.821\pm 1.057$&    $ 0.289\pm 0.022$\\
$h_1^{(4230)}/g_1^{(4230)}$   &   $ -1.022^{+0.081}_{-0.097}$ & $ -1.079\pm 0.016$&  $ 0.033\pm 0.333$&     $ -0.906\pm 0.024$\\
$h_8^{(4230)}/g_1^{(4230)}$    &   $ 1.186^{+0.144}_{-0.121}$&   $ 1.377\pm 0.025$&   $ -3.633\pm 1.122$&  $ 0.494\pm 0.038$ \\
$y^{(4230)}/g_1^{(4230)}~[\text{GeV}^{\frac{1}{2}}]$  &   $ 0.303^{+0.080}_{-0.120}$ & $2.640\pm 0.058$&   $ 1280\pm 347$& $353\pm 29$ \\
$g_{YP1}^{(4230)}/g_1^{(4230)}~[\text{GeV}^{\frac{1}{2}}]$     &   $ -113^{+10}_{-10}$ &   $ -149\pm 4$&   $ -60\pm 26$&   $ -124\pm 7$\\
$g_{YP2}^{(4230)}/g_1^{(4230)}~[\text{GeV}^{-\frac{1}{2}}]$     & $ -551^{+58}_{-50}$ &   $ -652\pm 27$&   $ 667\pm 102$&   $ 756\pm 76$\\
$g_{YP3}^{(4230)}/g_1^{(4230)}~[\text{GeV}^{-\frac{1}{2}}]$     &  $ 210^{+26}_{-39}$ &   $ 440\pm 26$&   $ 400\pm 131$&   $ -522\pm 84$\\
$j_1^{(4230)}/g_1^{(4230)}~[\text{GeV}]$      &   $ 39.7^{+1.5}_{-1.6}$&   $ 24.8\pm 1.0$ &   $ -$&   $ -$\\
$j_3^{(4230)}/g_1^{(4230)}~[\text{GeV}]$     &   $ 19.7^{+2.8}_{-3.0}$  &   $ 10.6\pm 1.7$&   $ -$&   $ -$\\
$j_5^{(4230)}/g_1^{(4230)}~[\text{GeV}]$      &   $ -4.8^{+3.2}_{-2.4}$&   $ -8.3\pm 2.6$&   $ -$&   $ -$ \\
$g_1^{(4260)}\times 10^{3}~[\text{GeV}^{-1}]$     &   $ 6.5^{+0.9}_{-0.9}$ &   $ 8.8\pm 0.1$  &   $ 3.8\pm 0.1$& $ 6.1\pm 0.1$ \\
$g_8^{(4260)}/g_1^{(4260)}$     &   $ 0.052^{+0.008}_{-0.006}$& $ 0.023\pm 0.006$& $ 1.139\pm 0.009$&   $ 0.602\pm 0.010$\\
$h_1^{(4260)}/g_1^{(4260)}$   &   $ -0.892^{+0.006}_{-0.006}$ &   $ -0.939\pm 0.004$&   $ -0.609\pm 0.007$&  $ -0.809\pm 0.002$  \\
$h_8^{(4260)}/g_1^{(4260)}$    &   $ 0.621^{+0.011}_{-0.013}$&   $ 0.795\pm 0.008$&   $ -0.714\pm 0.013$&  $ 0.136\pm 0.003$ \\
$y^{(4260)}/g_1^{(4260)}~[\text{GeV}^{\frac{1}{2}}]$   &   $ 0.524^{+0.216}_{-0.247}$& $3.550\pm 0.019$&   $ 654\pm 16$&$108\pm 21$ \\
$g_{YP1}^{(4260)}/g_1^{(4260)}~[\text{GeV}^{\frac{1}{2}}]$    &   $ -145^{+10}_{-9}$&   $ -237\pm 3$&   $ 42\pm 6$&  $ -233\pm 9$   \\
$g_{YP2}^{(4260)}/g_1^{(4260)}~[\text{GeV}^{-\frac{1}{2}}]$    &   $ -507^{+28}_{-18}$&   $ -662\pm 17$&   $ 290\pm 2$&  $ 84\pm 3$   \\
$g_{YP3}^{(4260)}/g_1^{(4260)}~[\text{GeV}^{-\frac{1}{2}}]$    &   $ 145^{+21}_{-19}$&   $ 345\pm 17$&   $ 187\pm 5$&  $ 67\pm 3$   \\
$j_1^{(4260)}/g_1^{(4260)}~[\text{GeV}]$      &   $ -33.6^{+6.2}_{-5.4}$&   $ -22.2\pm 2.2$&   $ -$&   $ -$ \\
$j_3^{(4260)}/g_1^{(4260)}~[\text{GeV}]$      &   $ 11.1^{+4.0}_{-4.3}$&   $ 1.1\pm 4.5$&   $ -$&   $ -$ \\
$j_5^{(4260)}/g_1^{(4260)}~[\text{GeV}]$      &   $ 29.3^{+9.2}_{-10.8}$&   $ 26.6\pm 3.1$&   $ -$&   $ -$ \\
$C_{12} ~[\text{fm}^{2}]$    &   $-0.006^{+0.006}_{-0.013}$&   $ -0.002\pm 0.001$&   $ -0.007\pm 0.002$&   $ -0.003\pm 0.001$ \\
$C_{1Z}~[\text{fm}^{2}]$    &   $ -0.208^{+0.004}_{-0.004}$ &   $ -0.227\pm 0.002$&   $ -0.212\pm 0.001$&   $ -0.230\pm 0.002$ \\
$b~[\text{fm}^{3}]$   &  $ -0.338^{+0.043}_{-0.054}$  &   $ 0$ $(\text{fixed})$ &   $ -0.187\pm 0.005$&   $ 0$ (fixed)\\
\hline
 $\frac{{\chi^2}}{{\rm \text{number of data points}}}$ of (4230, 4260)   &  $(\frac{307.3}{152},\frac{164.5}{160})$ &  $(\frac{359.9}{152},\frac{262.8}{160})$ &  $(\frac{403.8}{152},\frac{173.3}{160})$&  $(\frac{444.1}{152},\frac{287.7}{160})$\\
\end{tabular}
\end{ruledtabular}
\end{table}

\begin{table}[tbh]
\caption{\label{tablepar_T} Correlation coefficients in the $J/\psi\pi$--$D\bar{D}^{\ast}$ coupled-channel $T$-matrix in Fit I. }
\begin{center}
\renewcommand{\arraystretch}{1.35}
\begin{tabular}{cccc}
\toprule
      & $C_{12}$ & $C_{1Z}$ & $b$  \\
\hline
$C_{12}$    &   $1$&   $0.19$&   $0.19$ \\
$C_{1Z}$    &   $0.19$ &   $1$&   $0.99$ \\
$b$   &     $0.19$  &   $0.99$ &   $1$\\
\botrule
\end{tabular}
\end{center}
\end{table}

\end{onecolumngrid}

\end{document}